\documentclass[aps,pra,showpacs,twocolumn]{revtex4-1}
\usepackage{amssymb}
\usepackage{amsmath}
\usepackage{graphicx}
\usepackage{epsfig}

\begin{document}

\title{Finite-temperature quantum criticality in a complex-parameter plane}
\author{C. Li and Z. Song}
\email{songtc@nankai.edu.cn}
\affiliation{School of Physics, Nankai University, Tianjin 300071, China}

\begin{abstract}
A conventional quantum phase transition (QPT) occurs not only at zero
temperature, but also exhibits finite-temperature quantum criticality.
Motivated by the discovery of the pseudo-Hermiticity of non-Hermitian
systems, we explore the finite-temperature quantum criticality in a
non-Hermitian $\mathcal{PT}$-symmetric Ising model. We present the complete
set of exact eigenstates of the non-Hermitian Hamiltonian, based on which
the mixed-state fidelity in the context of biorthogonal bases is calculated.
Analytical and numerical results show that the fidelity approach to
finite-temperature QPT can be extended to the non-Hermitian Ising model.
This paves the way for experimental detection of quantum criticality in a
complex-parameter plane.
\end{abstract}

\maketitle


\section{Introduction}

\label{sec_intro}

A central task of the theory of quantum phase transitions (QPTs) is to
describe the consequences of this zero temperature singularity on physical
properties at finite temperature since all experiments can only be done at
non-zero temperature \cite{S.Sachdev}. It turns out that the mixed-state
fidelity, which is related to the statistical distance between two density
operators,\ is a powerful way to describe the signatures of QPTs at nonzero
temperature \cite{Sun}.

Recently, there have been intense efforts to establish a $\mathcal{PT}$%
-symmetric quantum theory as a complex extension of the conventional quantum
mechanics \cite{Bender 98,Bender 99,Dorey 01,Bender
02,A.M43,A.M,A.M36,Jones,AM}, since the discovery that a non-Hermitian
Hamiltonian having simultaneous parity-time ($\mathcal{PT}$) symmetry has a
real spectrum \cite{Bender 98}. Motivated by the pseudo-Hermiticity of
non-hermitian systems, the QPT in a non-Hermitian $\mathcal{PT}$-symmetric
Ising model, which is driven by a complex staggered transverse field, has
been explored based on the exact solution \cite{Li}. It has been shown that
the phase diagram can be characterized by the geometric phase. The phase
boundary can be identified by a divergence of Berry curvature density, which
is defined in the context of biorthogonal bases.

In this paper, we address two questions concerning the quantum criticality
for the non-hermitian Ising model at finite temperature. At first, whether
the QPT driven by complex parameters has residual criticality at finite
temperature. Second, as a metric-based approach, whether the mixed-state
fidelity in the context of biorthogonal bases is available to describe this
criticality. Analytical and numerical results indicate that the phase
boundary at finite temperature can be identified by the mixed-state fidelity.

The paper is organized as follows: In Sec. \ref{sec_model} we present the
exact solution of the non-Hermitian Ising model. In Sec. \ref{sec_fidelity},
the mixed-state fidelity of thermal states is obtained analytically and a
numerical simulation is performed to demonstrate the signatures of QPTs at
finite temperature. Section \ref{sec_summary} contains the conclusion.

\begin{figure}[tbp]
\includegraphics[ bb=68 374 526 775, width=0.45\textwidth, clip]{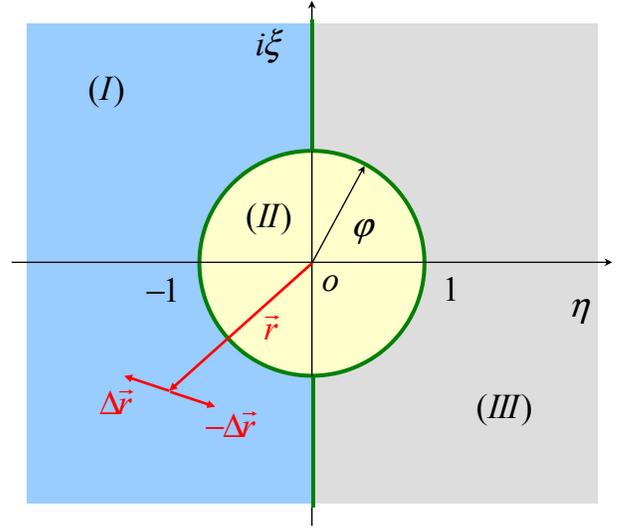}
\caption{(Color online) Phase diagram for the ground state of the Ising ring
in a complex staggered transverse field. The heavy lines (green) represent
the boundary which separates three quantum phases. Phases I and III are
paramagnet, while II is ferromagnet. Vector $\protect\overrightarrow{r}$\
denotes an arbitrary position in the $\protect\xi \protect\eta $\ plane,
while $\protect\overrightarrow{r}\pm \Delta \protect\overrightarrow{r}$ are
two points in the vicinity of point $\protect\overrightarrow{r}$. The
mixed-state fidelity for those two points can characterize the thermal
criticality at point $\protect\overrightarrow{r}$. } \label{fig1}
\end{figure}

\section{Model and solution}

\label{sec_model}

In the last few years, there have been many studies of the phase diagrams
for various quantum models in the aid of the fidelity \cite%
{Zanardi,Vezzani,Chen,Yang,Huan,Alet,Rams} or mixed-state fidelity \cite%
{Sun,Vieira,Quan,Sirker}. All the investigations are performed for Hermitian
Hamiltonians, where the probability is preserving in the context of standard
Dirac inner product. It would be interesting to explore the extension of
this metric-based approach to a non-Hermitian Hamiltonian with
pseudo-Hermiticity, where biorthonormal inner product is introduced. We
begin our investigation by introducing a non-Hermitian Ising model, whose
ground state phase diagram has been studied in previous work \cite{Li}. It
has been shown that the Berry phase in the context of biorthonormal inner
product can identify the phase diagram, which sheds some light on the
extension of the metric-based approach to a non-Hermitian system.

The concept is studied exemplarily for criticality in a non-Hermitian
one-dimensional spin-$1/2$ Ising model with a complex staggered\ transverse
magnetic field on a $2N$-site lattice. The system is modeled by the
following Hamiltonian%
\begin{equation}
\mathcal{H}\text{ }=-J\underset{j=1}{\overset{2N}{\sum }}\left( \sigma
_{j}^{z}\sigma _{j+1}^{z}+g_{_{j}}\sigma _{j}^{x}\right) ,  \label{H_general}
\end{equation}%
where complex field $g_{_{j}}=\eta +i\left( -1\right) ^{j}\xi $\ ($i=\sqrt{-1%
}$) with $\eta $ and $\xi $\ being real numbers. Here $\sigma _{j}^{\lambda
} $ ($\lambda =x,$ $z$) are the Pauli operators on site $j$, and satisfy the
periodic boundary condition $\sigma _{j}^{\lambda }\equiv \sigma
_{j+2N}^{\lambda }$. In ground state this model exhibits quantum critical
points at $\eta ^{2}+\xi ^{2}=1$ and $\eta =0$\ for $\eta ^{2}+\xi ^{2}>1$,
separating paramagnetic and ferromagnetic phases \cite{Li}.

Following the derivation in Appendix, all the eigenstates of $\mathcal{H}$\
can be constructed as the form
\begin{equation}
\left\vert \Psi \right\rangle =\prod_{k\in \left[ 0,\pi \right] }\overline{%
\left\vert n\right\rangle }_{k},
\end{equation}%
with eigen energy
\begin{equation}
E=\sum_{\left\{ \overline{\left\vert n\right\rangle }_{k}\right\} }\epsilon
^{n}\left( k\right) .
\end{equation}%
Here $\overline{\left\vert n\right\rangle }_{k}$ $\left( n\in \left[ 1,16%
\right] \right) $\ denotes the $16$-dimensional qudit ($16$-level system)
state, which is expressed explicitly as

\begin{eqnarray}
&&n\in \left[ 1,5\right] :\overline{\left\vert n\right\rangle }_{k}  \notag
\\
&=&\frac{1}{\Omega ^{n}}[2\cos \left( k/2\right) \left( \frac{\alpha
_{k}^{\dagger }\alpha _{-k}^{\dagger }}{\epsilon ^{n}+i2\xi }+\frac{\beta
_{k}^{\dagger }\beta _{-k}^{\dagger }}{\epsilon ^{n}-i2\xi }\right)  \notag
\\
&&-2i\sin \left( k/2\right) \left( \frac{1}{\epsilon ^{n}+2\eta }\alpha
_{k}^{\dagger }\beta _{k}^{\dagger }\alpha _{-k}^{\dagger }\beta
_{-k}^{\dagger }+\frac{1}{\epsilon ^{n}-2\eta }\right)  \notag \\
&&+e^{ik/2}\alpha _{k}^{\dagger }\beta _{-k}^{\dagger }-e^{-ik/2}\beta
_{k}^{\dagger }\alpha _{-k}^{\dagger }]\left\vert 0\right\rangle ,
\end{eqnarray}%
and%
\begin{eqnarray}
\overline{\left\vert 6\right\rangle }_{k} &=&\frac{1}{\sqrt{2}}\left(
e^{ik/2}\alpha _{k}^{\dagger }\beta _{-k}^{\dagger }+e^{-ik/2}\beta
_{k}^{\dagger }\alpha _{-k}^{\dagger }\right) \left\vert 0\right\rangle ,
\notag \\
\overline{\left\vert \overline{7}\right\rangle }_{k} &=&\alpha _{k}^{\dagger
}\beta _{k}^{\dagger }\left\vert 0\right\rangle \text{, }\left\vert
\overline{8}_{k}\right\rangle =\alpha _{-k}^{\dagger }\beta _{-k}^{\dagger
}\left\vert 0\right\rangle ,
\end{eqnarray}%
and%
\begin{eqnarray}
n &\in &\left[ 9,12\right] :\overline{\left\vert n\right\rangle }_{k}  \notag
\\
&=&\frac{1}{\Omega ^{n}}\{\{\left( \epsilon ^{n}+\eta -i\xi \right) [\left(
\epsilon ^{n}\right) ^{2}-\left( \eta +i\xi \right) ^{2}-1]  \notag \\
&&-2(\eta \cos ^{2}\frac{k}{2}-i\xi \sin ^{2}\frac{k}{2})\}e^{-ik/2}\alpha
_{k}^{\dagger }  \notag \\
&&+[\left( \epsilon ^{n}+\eta \right) ^{2}+\xi ^{2}-1]\cos \frac{k}{2}\beta
_{k}^{\dagger }  \notag \\
&&-i[\left( \epsilon ^{n}-i\xi \right) ^{2}-\eta ^{2}-1]\sin \frac{k}{2}%
\alpha _{-k}^{\dagger }\alpha _{k}^{\dagger }\beta _{k}^{\dagger }  \notag \\
&&-i\left( \eta +i\xi \right) \sin ke^{-ik/2}\alpha _{k}^{\dagger }\beta
_{-k}^{\dagger }\beta _{k}^{\dagger }\}\left\vert 0\right\rangle ,
\end{eqnarray}%
and%
\begin{eqnarray}
n &\in &\left[ 12,16\right] :\overline{\left\vert n\right\rangle }_{k}
\notag \\
&=&\frac{1}{\Omega ^{n}}\{\{\left( \epsilon ^{n}+\eta -i\xi \right) [\left(
\epsilon ^{n}\right) ^{2}-\left( \eta +i\xi \right) ^{2}-1]  \notag \\
&&-2(\eta \cos ^{2}\frac{k}{2}-i\xi \sin ^{2}\frac{k}{2})\}e^{-ik/2}\alpha
_{-k}^{\dagger }  \notag \\
&&+[\left( \epsilon ^{n}+\eta \right) ^{2}+\xi ^{2}-1]\cos \frac{k}{2}\beta
_{-k}^{\dagger }  \notag \\
&&-i[\left( \epsilon ^{n}-i\xi \right) ^{2}-\eta ^{2}-1]\sin \frac{k}{2}%
\alpha _{k}^{\dagger }\alpha _{-k}^{\dagger }\beta _{-k}^{\dagger }  \notag
\\
&&-i\left( \eta +i\xi \right) \sin ke^{-ik/2}\alpha _{-k}^{\dagger }\beta
_{k}^{\dagger }\beta _{-k}^{\dagger }\}\left\vert 0\right\rangle ,
\end{eqnarray}%
where $\Omega ^{n}$\ is corresponding normalization factor. And coefficients
$\epsilon ^{n}\left( k\right) $\ are given as%
\begin{eqnarray}
\epsilon ^{1} &=&[2r^{2}\cos \left( 2\varphi \right) +2+2\sqrt{%
r^{4}-2r^{2}\cos k+1}]^{1/2},  \notag \\
\epsilon ^{3} &=&[2r^{2}\cos \left( 2\varphi \right) +2-2\sqrt{%
r^{4}-2r^{2}\cos k+1}]^{1/2},  \notag \\
\epsilon ^{2} &=&-\epsilon ^{1},\epsilon ^{4}=-\epsilon ^{3},\text{ }%
\epsilon ^{5}=\epsilon ^{6}=\epsilon ^{7}=\epsilon ^{8}=0,  \notag \\
\epsilon ^{9} &=&[r^{2}\cos \left( 2\varphi \right) +1  \notag \\
&&+\sqrt{4r^{2}\cos ^{2}\varphi +2r^{2}\cos k-2r^{4}}]^{1/2},  \notag \\
\epsilon ^{11} &=&[r^{2}\cos \left( 2\varphi \right) +1  \notag \\
&&-\sqrt{4r^{2}\cos ^{2}\varphi +2r^{2}\cos k-2r^{4}}]^{1/2},  \notag \\
\epsilon ^{14} &=&-\epsilon ^{13}=\epsilon ^{10}=-\epsilon ^{9},\text{ }
\notag \\
\epsilon ^{16} &=&-\epsilon ^{15}=\epsilon ^{12}=-\epsilon ^{11}.
\end{eqnarray}%
Here\ $\alpha _{k}^{\dagger }$ and $\beta _{k}^{\dagger }$ are fermionic
creation operators and we parameterize the complex field in terms of the
polar radius and angle

\begin{equation}
r=\sqrt{\eta ^{2}+\xi ^{2}}\text{ and }\tan \varphi =\xi /\eta .
\end{equation}%
Similarly, we can obtain the eigenstates of $\mathcal{H}^{\dag }$ by the
qudit state $\left\langle n\right\vert _{k}$, which has the form

\begin{equation}
\left\langle n\left( \xi \right) \right\vert _{k}=\left( \overline{%
\left\vert n\left( -\xi \right) \right\rangle }_{k}\right) ^{\dag }.
\end{equation}%
It is turned out that the complete biorthonormal set can be constructed due
to the fact%
\begin{equation}
\left\langle n\right\vert _{k}\overline{\left\vert m\right\rangle }%
_{k^{\prime }}=\delta _{mn}\delta _{kk^{\prime }}\text{, }\sum_{n}\overline{%
\left\vert n\right\rangle }_{k}\left\langle n\right\vert _{k}=1.  \label{bio}
\end{equation}

\begin{figure*}[tbp]
\includegraphics[ bb=33 20 542 533, width=0.32\textwidth, clip]{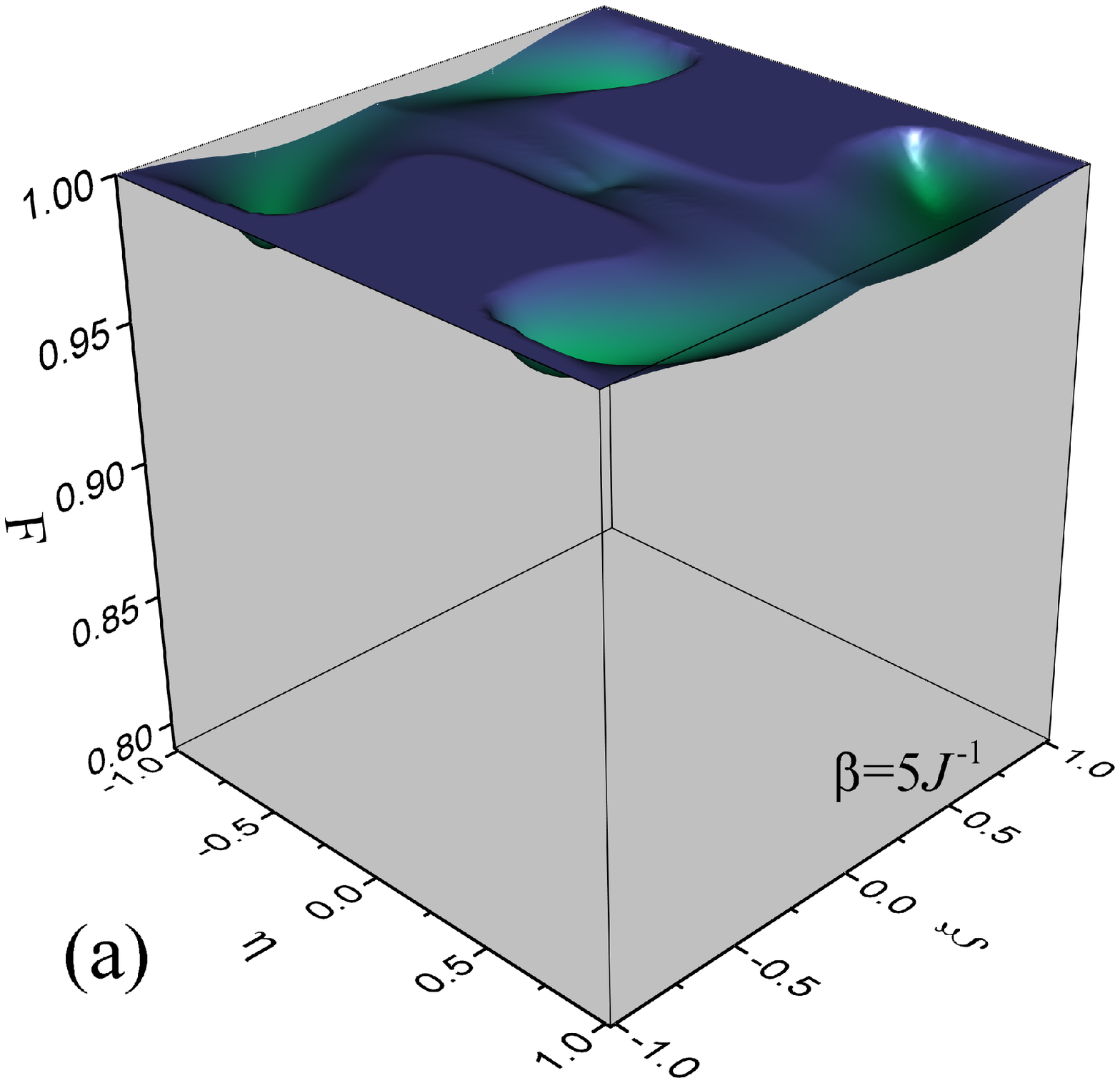} %
\includegraphics[ bb=33 20 542 533, width=0.32\textwidth, clip]{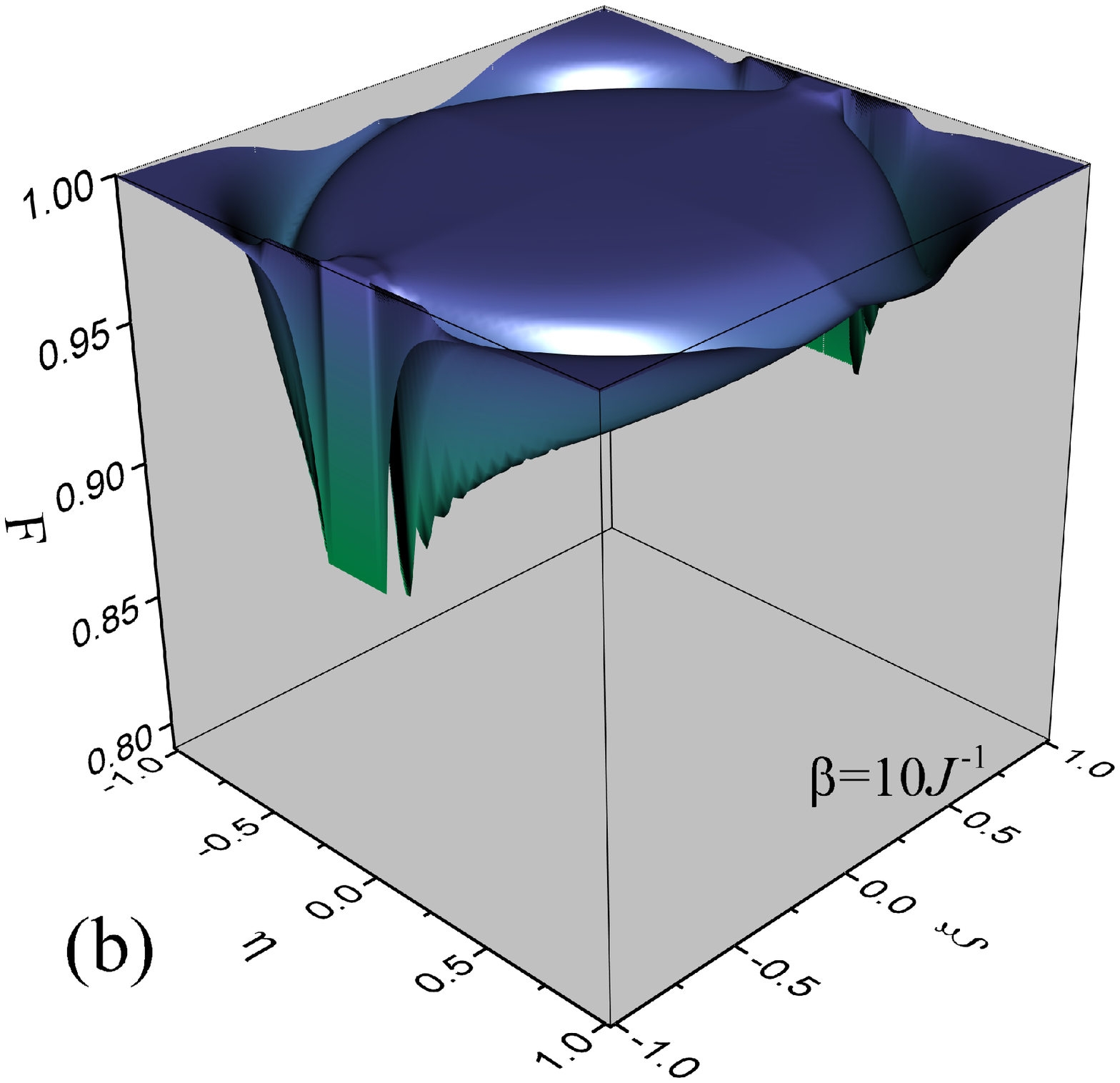} %
\includegraphics[ bb=33 20 542 533, width=0.32\textwidth, clip]{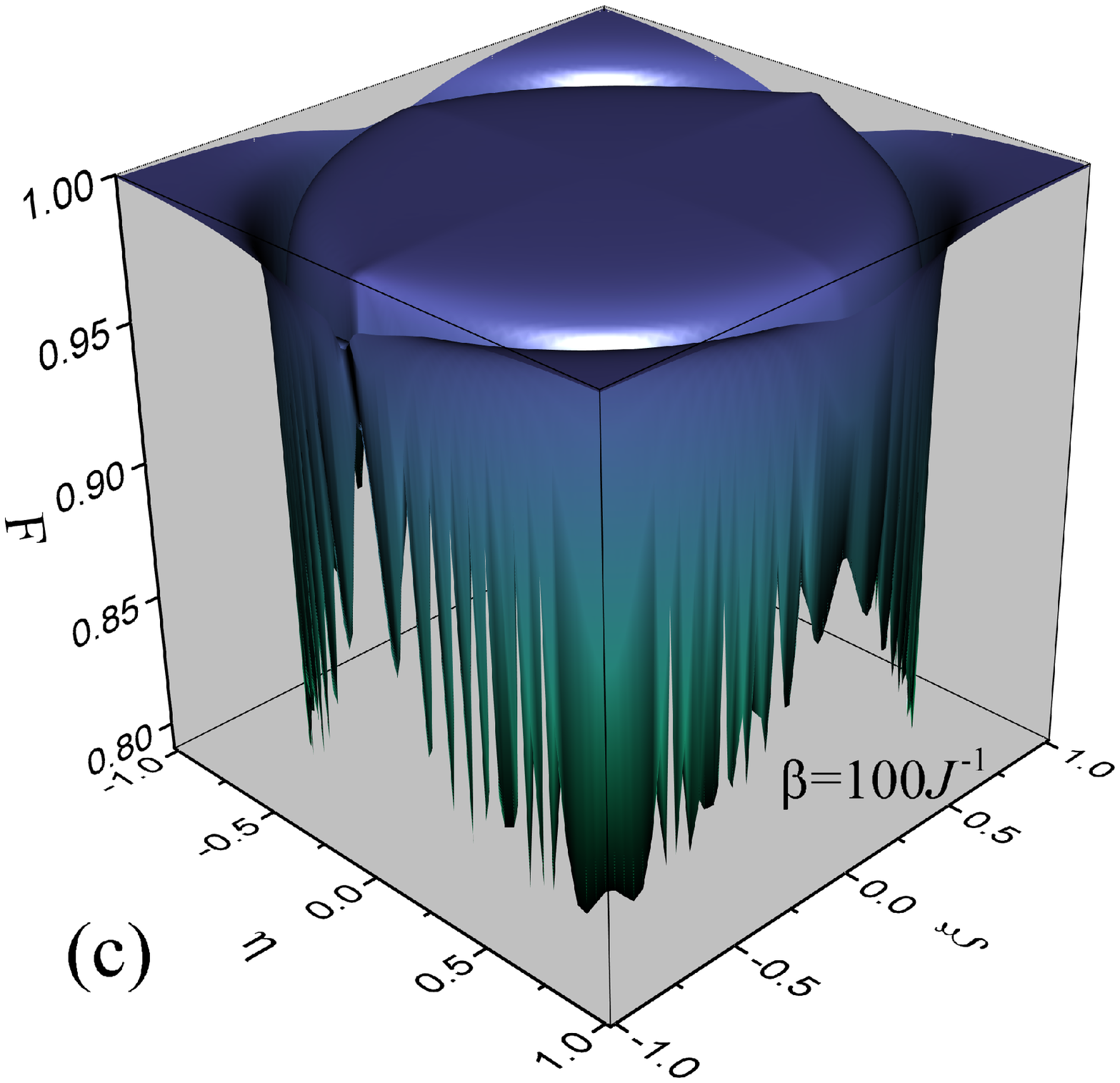} %
\includegraphics[ bb=33 20 542 533, width=0.32\textwidth, clip]{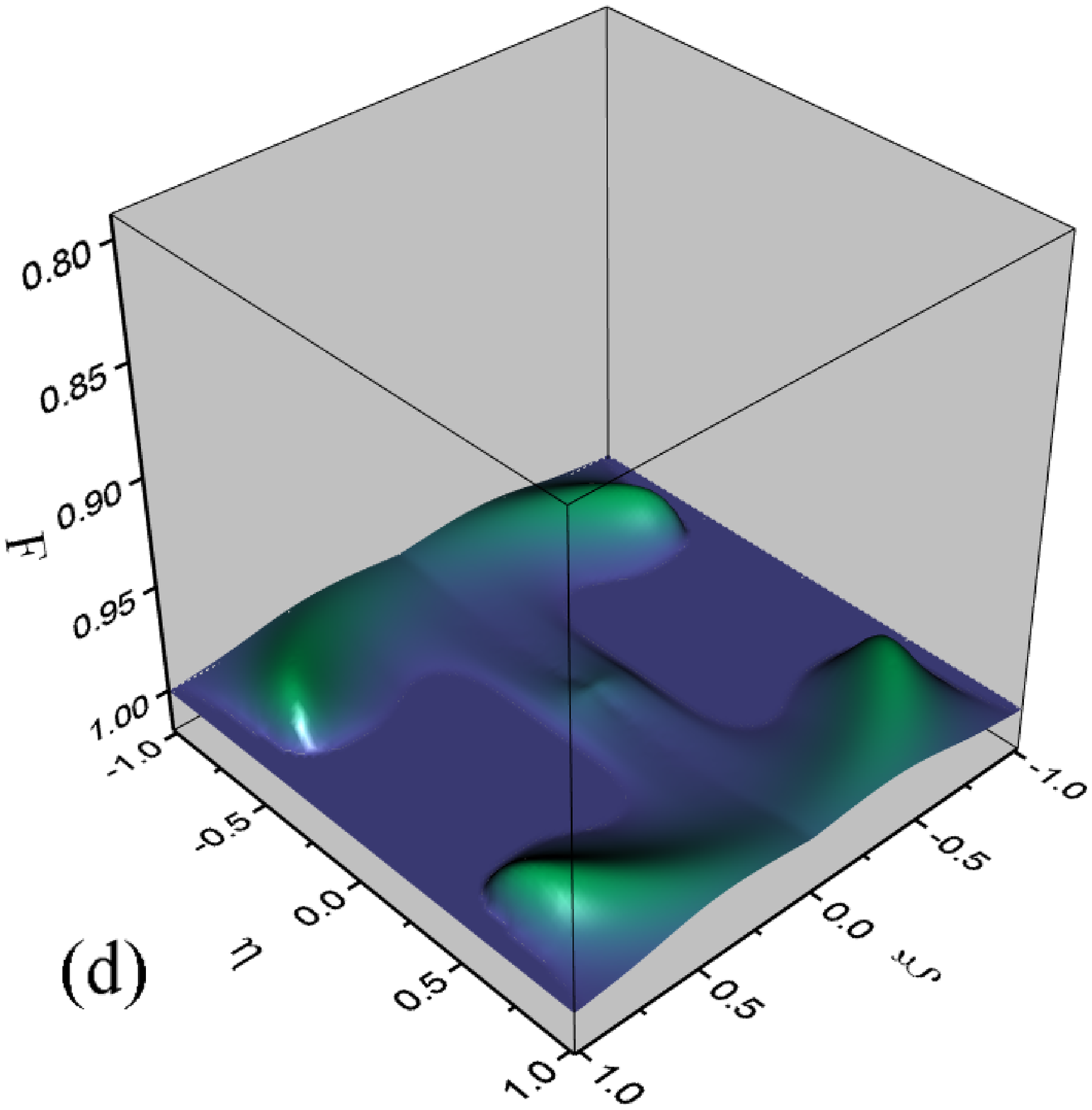} %
\includegraphics[ bb=33 20 542 533, width=0.32\textwidth, clip]{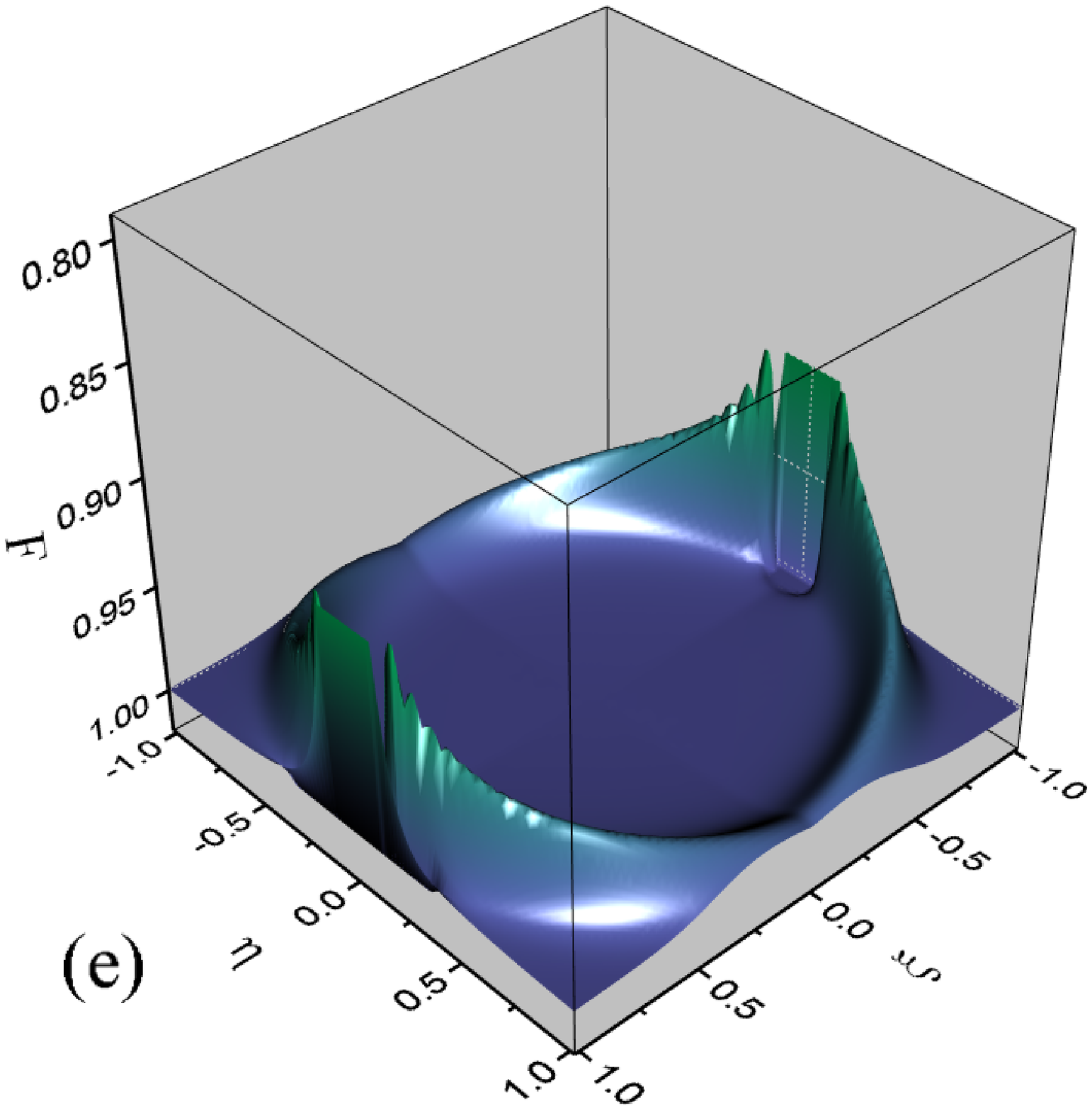} %
\includegraphics[ bb=33 20 542 533, width=0.32\textwidth, clip]{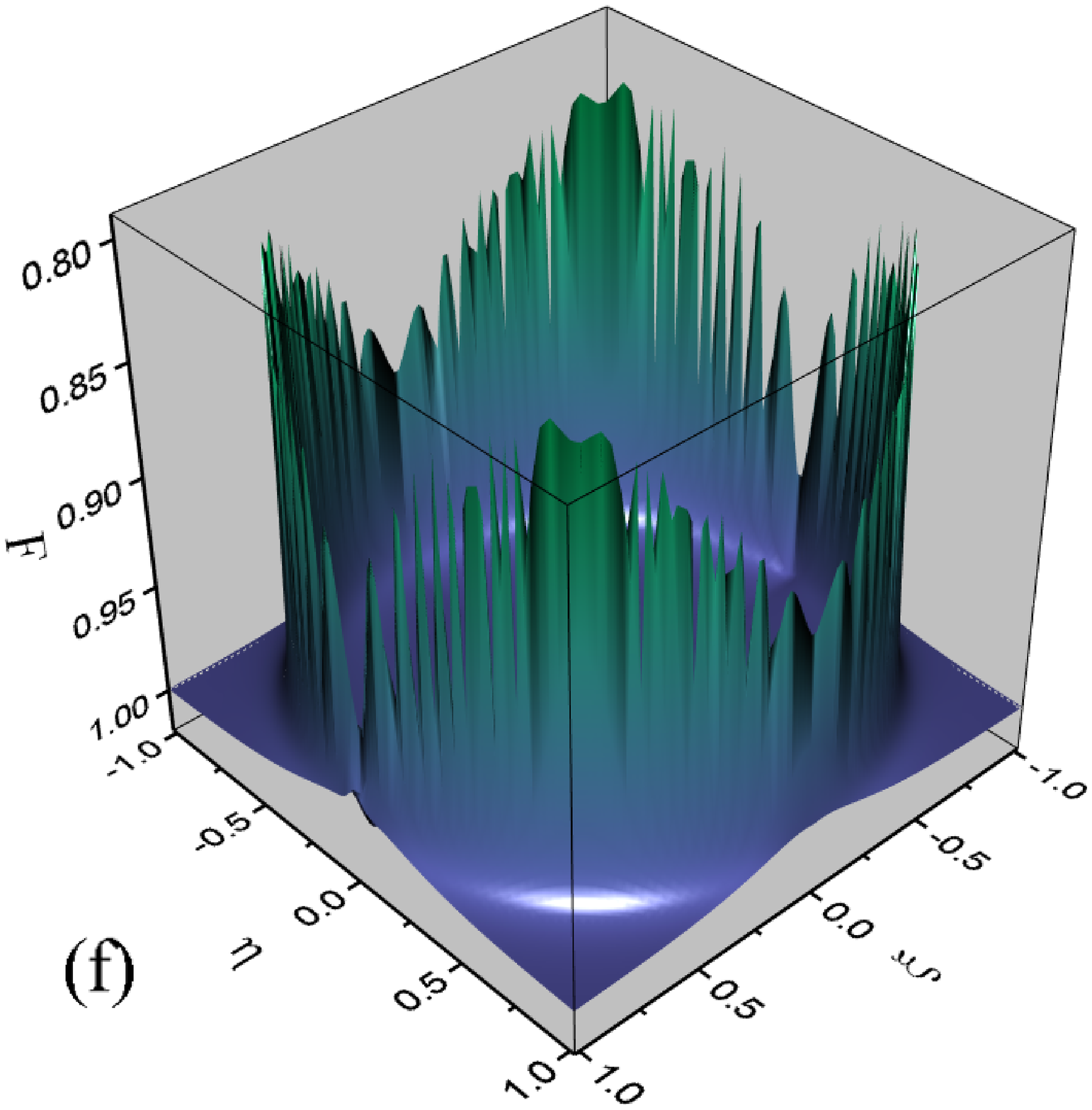}
\caption{(Color online) $3$-D plots of mixed-state fidelity of
finite-temperature thermal state of the non-Hermitian Ising model. We
consider the case with $N=300$ and the field variances $\Delta \protect\eta %
=\Delta \protect\xi =0.01J^{-1}$. Figures (a-c) plots of $F$ as functions of
$\protect\eta $ and $\protect\xi $\ for different temperatures $\protect%
\beta $, while (d-f)\ are corresponding inversions of\ (a-c). At high
temperatures (a), the mixed-state fidelity shows a flat surface. When the
temperature decreases (b, c), the decay of fidelity becomes sharp and
clearly shows the signature of QPT which occurs at absolute zero
temperature. A more accurate dependence is plotted in Fig.3.}
\label{fig2}
\end{figure*}

At zero-temperature, the phase diagram is obtained in Ref. \cite{Li} and
sketched in Fig.\ref{fig1}. The influence of the imaginary field $i\xi $\ on
the phase transition is obvious: It enhances the action of the real
transverse field $\eta $, shrinking the disorder region, leading to a
circular phase boundary. In the following, we will show that this circle
critical line is still robust at finite temperature.

\section{Mixed-state fidelity}

\label{sec_fidelity}The key point of this paper is that the mixed-state
fidelity%
\begin{equation}
F\left( \rho ,\widetilde{\rho }\right) =\mathrm{tr}\sqrt{\widetilde{\rho }%
^{1/2}\rho \widetilde{\rho }^{1/2}}  \label{F}
\end{equation}%
can be utilized to characterize the thermal criticality although density
matrices $\rho $ and $\widetilde{\rho }$ are non-Hermitian. The
non-Hermiticity of the matrices arises from the description of two different
thermal states

\begin{eqnarray}
\rho &=&\exp \left( -\beta \mathcal{H}\right) /\mathrm{tr}\exp \left( -\beta
\mathcal{H}\right) , \\
\widetilde{\rho } &=&\exp \left( -\beta \widetilde{\mathcal{H}}\right) /%
\mathrm{tr}\exp \left( -\beta \widetilde{\mathcal{H}}\right) ,
\end{eqnarray}%
where%
\begin{equation}
\mathcal{H}=\mathcal{H}\left( \eta ,\xi \right) ,\widetilde{\mathcal{H}}=%
\mathcal{H}\left( \eta +\delta \eta ,\xi +\delta \eta \right) ,  \label{HH}
\end{equation}%
are both non-Hermitian operators.

As the consequence of the translational symmetry, $\left[ h_{k},h_{k^{^{%
\prime }}}\right] =0$, leads to%
\begin{equation}
F\left( \rho ,\widetilde{\rho }\right) =\prod_{k}\mathrm{tr}\sqrt{\rho
_{k}^{1/2}\widetilde{\rho }_{k}\rho _{k}^{1/2}},
\end{equation}%
which can simplify the computation procedure. Studies on Hermitian systems
show that a zero-temperature fidelity vanishes at the quantum boundary. It
is expected that the similar behavior can occur for the present
non-Hermitian model. Actually,\ in the limit case $\beta \rightarrow \infty $%
, $F\left( \rho ,\widetilde{\rho }\right) $\ is reduced to%
\begin{equation}
F_{\infty }\left( \rho ,\widetilde{\rho }\right)
=\prod_{k}O_{k}=\prod_{k}\left\vert \widetilde{\left\langle 1\right\vert }%
_{k}\overline{\left\vert 1\right\rangle }_{k}\right\vert ,
\end{equation}%
by using the completeness condition in Eq. (\ref{bio}). Here states\textbf{\
}$\widetilde{\left\langle 1\right\vert }_{k}$\textbf{\ }and $\overline{%
\left\vert 1\right\rangle }_{k}$\textbf{\ }are%
\begin{equation}
\widetilde{\left\langle 1\right\vert }_{k}=\left\langle 1\left(
\overrightarrow{r}+\Delta \overrightarrow{r}\right) \right\vert _{k},%
\overline{\left\vert 1\right\rangle }_{k}=\overline{\left\vert 1\left(
\overrightarrow{r}-\Delta \overrightarrow{r}\right) \right\rangle }_{k},
\end{equation}%
where $\overrightarrow{r}$\ denotes a vector in the $\xi \eta $\ plane (see
Fig.\ref{fig1}). In the following,\textbf{\ }we consider a simple case with $%
\overrightarrow{r}=$ $\left( \cos \varphi ,\sin \varphi \right) $\ and $%
\Delta \overrightarrow{r}=$ $\left( \Delta r\cos \varphi ,\Delta r\sin
\varphi \right) $.\ Correspondingly, states\textbf{\ }$\widetilde{%
\left\langle 1\right\vert }_{k}$\textbf{\ }and\textbf{\ }$\overline{%
\left\vert 1\right\rangle }_{k}$\textbf{\ }can be expressed in a polar
coordinate system\textbf{\ }$(r,\varphi )$%
\begin{equation}
\widetilde{\left\langle 1\right\vert }_{k}=\left\langle 1\left( 1+\Delta
r,\varphi \right) \right\vert _{k},\overline{\left\vert 1\right\rangle }_{k}=%
\overline{\left\vert 1\left( 1-\Delta r,\varphi \right) \right\rangle }_{k}.
\end{equation}%
We note that any vanishing overlaps $\widetilde{\left\langle 1\right\vert }%
_{k}\overline{\left\vert 1\right\rangle }_{k}$\ can result in the zero point
of fidelity $F_{\infty }\left( \rho ,\widetilde{\rho }\right) $. In the
thermodynamic limit $N\rightarrow \infty $, $k$\ becomes a continuous
variable. This ensures us to estimate the overlap $\lim_{k\rightarrow 0}%
\widetilde{\left\langle 1\right\vert }_{k}\overline{\left\vert
1\right\rangle }_{k}$. Direct derivation shows that%
\begin{eqnarray}
&&\lim_{k\rightarrow 0}\overline{\left\vert 1\left( r,\varphi \right)
\right\rangle }_{k}=\frac{1}{\gamma +\gamma ^{\ast }}(\alpha _{k}^{\dagger
}\beta _{-k}^{\dagger }-\beta _{k}^{\dagger }\alpha _{-k}^{\dagger }  \notag
\\
&&+\gamma ^{\ast }\alpha _{k}^{\dagger }\alpha _{-k}^{\dagger }+\gamma \beta
_{k}^{\dagger }\beta _{-k}^{\dagger })\left\vert 0\right\rangle \text{, }%
\left( r<1\right) ,
\end{eqnarray}%
and%
\begin{equation}
\lim_{k\rightarrow 0}\widetilde{\left\langle 1\left( r,\varphi \right)
\right\vert }_{k}=\left\langle 0\right\vert i\text{, }\left( r>1\right) ,
\end{equation}%
where $\varphi \in \left[ 0,\frac{\pi }{2}\right) \cup \left( \frac{3\pi }{2}%
,2\pi \right] $, and $\gamma =\sqrt{1-r^{2}\sin \varphi }+ir\sin \varphi $.
Meanwhile we have

\begin{eqnarray}
&&\lim_{k\rightarrow 0}\overline{\left\vert 1\left( r,\varphi \right)
\right\rangle }_{k}=\frac{1}{\gamma +\gamma ^{\ast }}(\alpha _{k}^{\dagger
}\beta _{-k}^{\dagger }-\beta _{k}^{\dagger }\alpha _{-k}^{\dagger }  \notag
\\
&&+\gamma ^{\ast }\alpha _{k}^{\dagger }\alpha _{-k}^{\dagger }+\gamma \beta
_{k}^{\dagger }\beta _{-k}^{\dagger })\left\vert 0\right\rangle \text{, }%
\left( r<1\right) ,
\end{eqnarray}%
and%
\begin{equation}
\lim_{k\rightarrow 0}\widetilde{\left\langle 1\left( r,\varphi \right)
\right\vert }_{k}=\left\langle 0\right\vert i\alpha _{k}\beta _{k}\alpha
_{-k}\beta _{-k}\text{, }\left( r>1\right) ,
\end{equation}%
where $\varphi \in \left( \frac{\pi }{2},\frac{3\pi }{2}\right) $. These
lead to the conclusion that

\begin{equation}
\lim_{k\rightarrow 0}\left\langle 1\left( 1+\Delta r,\varphi \right)
\right\vert _{k}\overline{\left\vert 1\left( 1-\Delta r,\varphi \right)
\right\rangle }_{k}=0,
\end{equation}%
for $\Delta r>0$\ and $\varphi \neq \frac{\pi }{2},\frac{3\pi }{2}$. The
essence of the results is the nonanalytical behavior of states\textbf{\ }$%
\lim_{k\rightarrow 0}\widetilde{\left\langle 1\right\vert }_{k}$\textbf{\ }%
and\textbf{\ }$\lim_{k\rightarrow 0}\overline{\left\vert 1\right\rangle }%
_{k} $\textbf{\ }at the critical points. This indicates that the fidelity
approach for QPT can be extended into the complex regime at zero temperature.

However, the overlap in both points\ $\varphi _{0}=\frac{\pi }{2}$ and $%
\frac{3\pi }{2}$ can be calculated analytically as%
\begin{eqnarray}
&&\lim_{k\rightarrow 0}\left\vert \left\langle 1\left( 1+\Delta r,\varphi
_{0}\right) \right\vert _{k}\overline{\left\vert 1\left( 1-\Delta r,\varphi
_{0}\right) \right\rangle }_{k}\right\vert   \notag \\
&=&\frac{\Delta r}{\left( 1+\Delta r\right) \sqrt{2\Delta r-\left( \Delta
r\right) ^{2}}},
\end{eqnarray}%
for $\Delta r>0$,\ which is reduced to $\sqrt{\Delta r/2}$\ for $\Delta r\ll
1$.\ In this sense, points $\left( 1,\varphi _{0}\right) $\ are a little
special compared to other points on the circle. This may be due to the fact
that points $\left( 1,\varphi _{0}\right) $\ are three-phase points. In this
paper, we focus on the critical points on the circle except these two points.

\begin{figure*}[tbp]
\includegraphics[ bb=21 247 395 518, width=0.32\textwidth, clip]{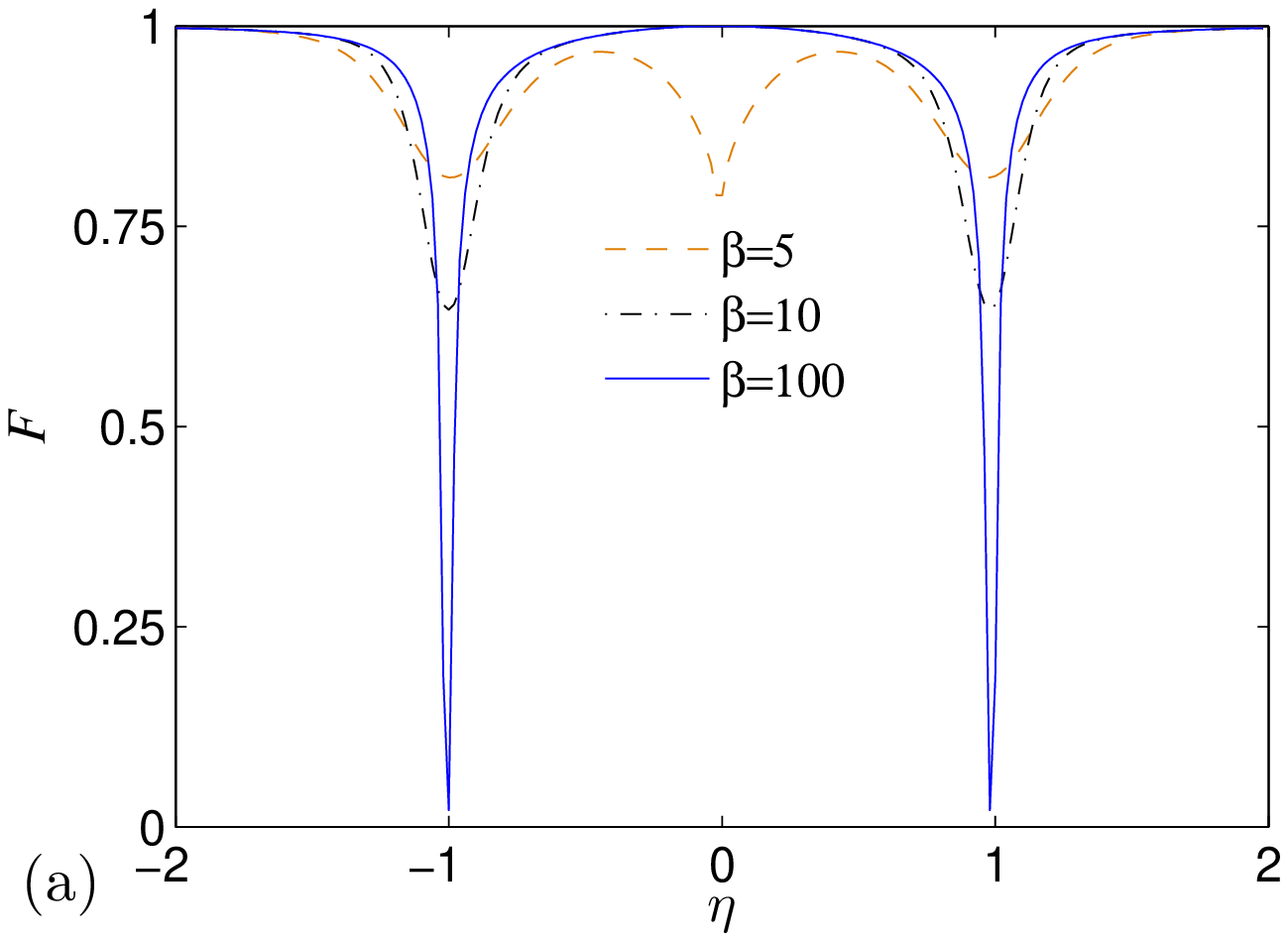} %
\includegraphics[ bb=21 247 395 518, width=0.32\textwidth, clip]{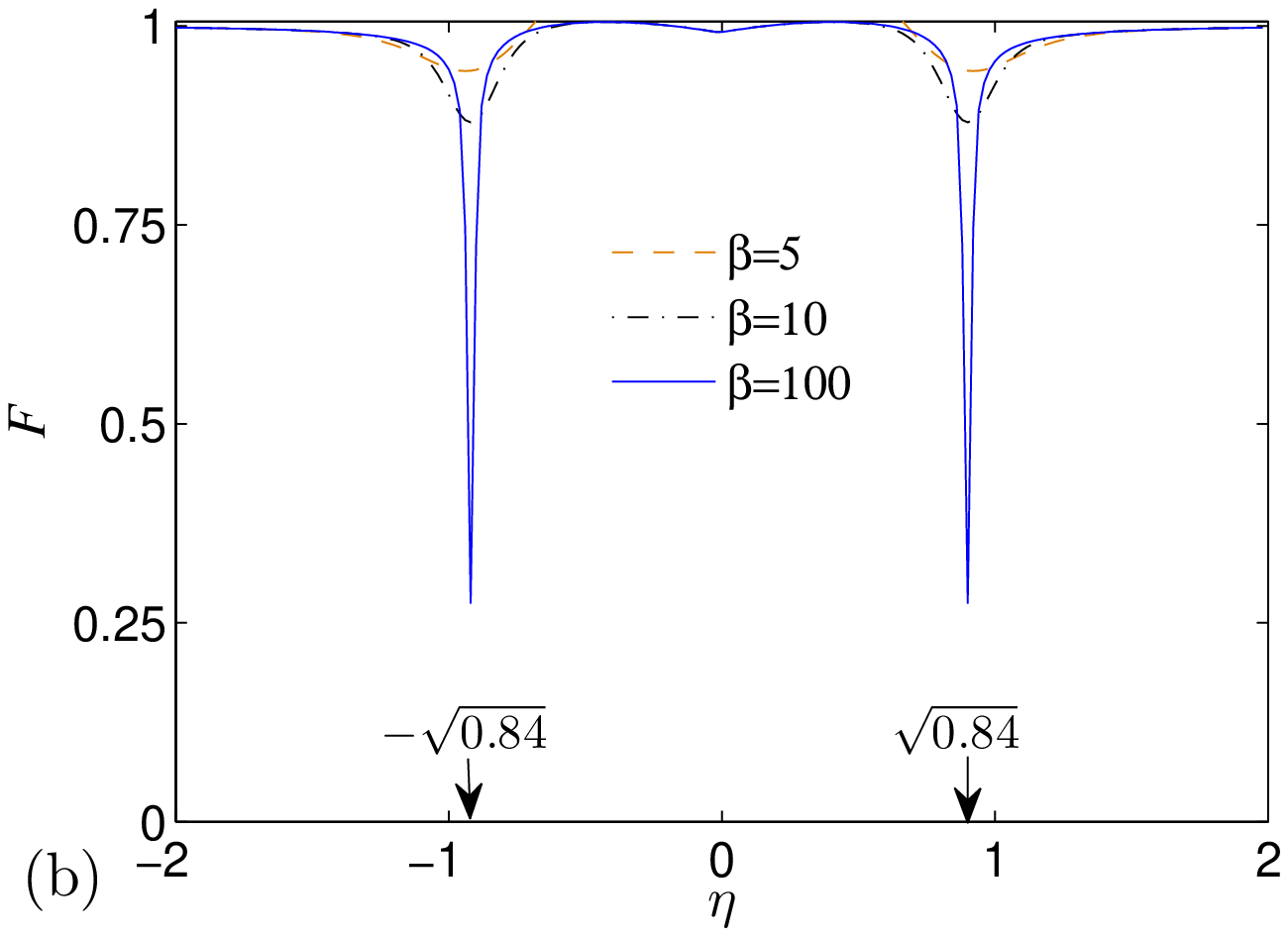} %
\includegraphics[ bb=21 247 395 518, width=0.32\textwidth, clip]{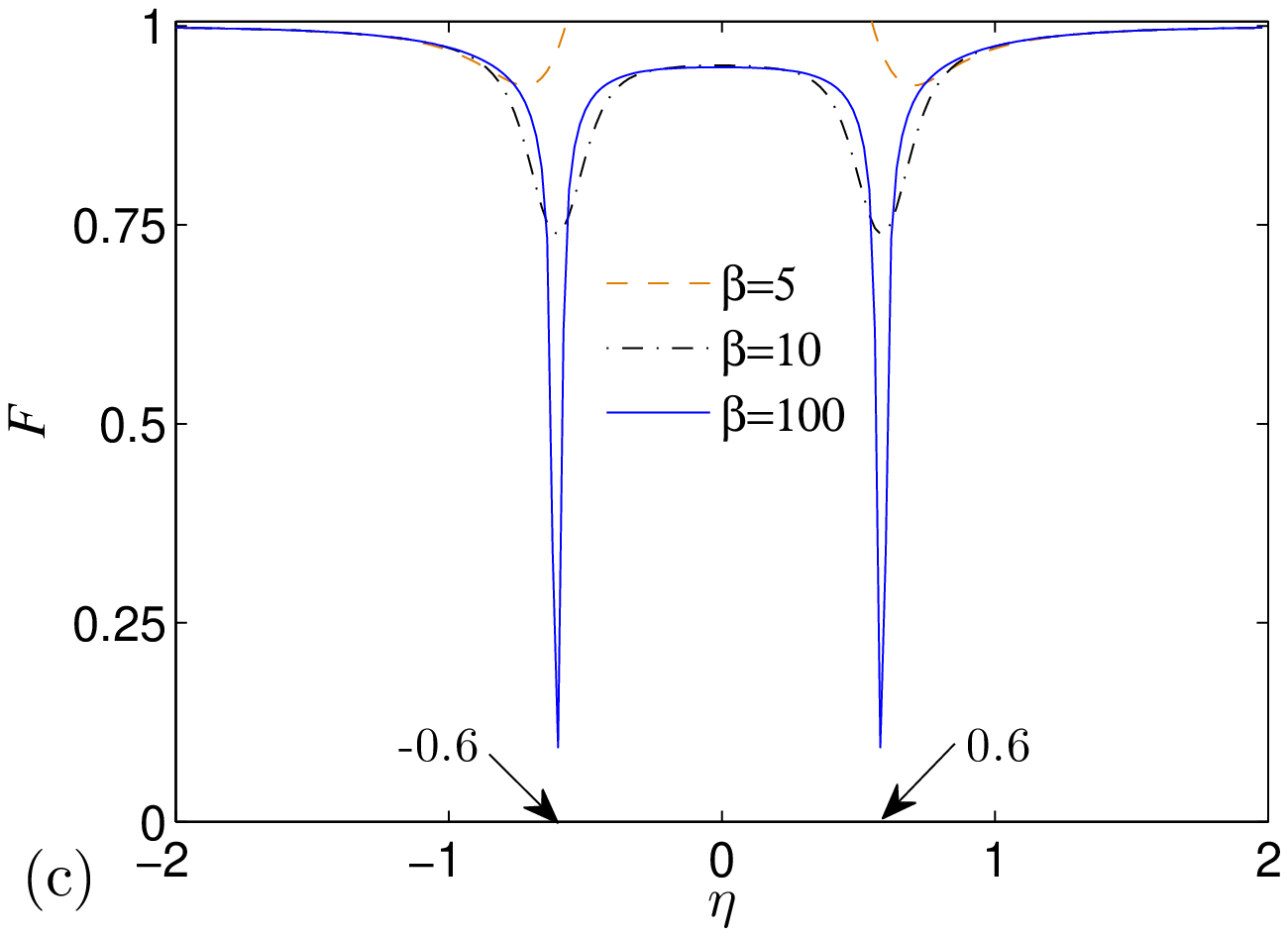}
\caption{(Color online) $2$-D plots of mixed-state fidelity of
finite-temperature thermal state of the non-Hermitian Ising model. We
consider the cases with $N=10^{4}$ and field parameters (a) $\protect\xi =0$%
, $\Delta \protect\eta =\Delta \protect\xi =0.01J^{-1}$, (b) $\protect\xi %
=0.4$, $\Delta \protect\eta =\Delta \protect\xi =0.01J^{-1}$ and (c) $%
\protect\xi =0.8$, $\Delta \protect\eta =\Delta \protect\xi =0.01J^{-1}$.
Three curves represent the fidelities at various temperatures indicated in
figures, respectively. It shows that the decay of fidelity for low
temperature well indicates the critical point of the non-Hermitian Ising
model. When the temperature increases, the decay of fidelity becomes less
sharp, leading to boundary broadening of QPT.}
\label{fig3}
\end{figure*}

\begin{figure}[tbp]
\includegraphics[ bb=83 218 447 567, width=0.45\textwidth, clip]{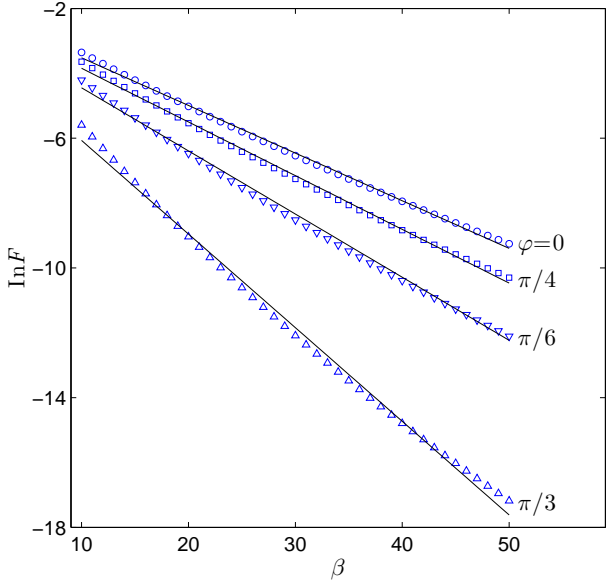}
\caption{(Color online) Plots of $F$\ as functions of temperature for
several typical values of $\protect\varphi $ and $N=10^{4}$, where $\beta$ is in the unit of $J^{-1}$. We fit the data
with the functions in Eqs. (\protect\ref{scaling1}) and (\protect\ref{scaling2}).}
\label{fig4}
\end{figure}

For finite temperature, numerical simulations are performed for different $%
\beta $. We plot $F$\ as a function of $\left( \xi ,\eta \right) $\ in Fig.%
\ref{fig2} and Fig.\ref{fig3}. We see that $F$\ is almost a constant and
equals to unity for a wide range of $\left( \xi ,\eta \right) $, apart from
the very narrow area around the circle $\eta ^{2}+\xi ^{2}=1$, where
drastically drops to zero as shown in Fig.\ref{fig3}. For larger $\beta $,
we can see that the dip of $F$\ well indicates the critical point of the
non-Hermitian Ising model for different values of $\xi $. We also study the
temperature dependence of the fidelity for different critical fields, which
is labeled\ by $\varphi $. In Fig. \ref{fig4} we plot $F$\ as functions of
temperature for several typical values of $\varphi $ and $N=10^{4}$. We fit
the data with the function
\begin{equation}
\ln F=\Gamma \beta +\ln A,  \label{scaling1}
\end{equation}%
which indicates the fidelity decays exponentially with decay constant $%
\Gamma $\ and the pre-exponential factor $A$. Here $\Gamma $\ and $A$\ are
functions of $\varphi $\ as the form%
\begin{eqnarray}
\Gamma  &=&\left[ -1.1+1.5\cos \left( 2\varphi \right) \right] \times 10^{5},
\notag \\
\ln A &=&\left[ -7.1+9.5\cos 2\varphi -2.4\cos \left( 4\varphi \right) %
\right] \times 10^{11}.  \label{scaling2}
\end{eqnarray}%
Based on above analytical and numerical results, we conclude that the
metric-based approach established for Hermitian systems can be applicable to
the non-Hermitian Ising model.

\section{Summary and discussion}

\label{sec_summary}

In summary, we have shown that the QPT in a non-Hermitian Ising model has
residual criticality at finite temperature, which can be described by the
mixed-state fidelity in the context of biorthogonal bases. The successful
extension of mixed-state fidelity to complex regime may be ascribed to the
following two facts: At first, we happen to have an exact solvable
non-Hermitian model, whose ground state is divided into several regions, or
phases in the a complex-parameter plane. So far, this model is unique, while
the QPT in most discovered non-Hermitian models is referred as the
exceptional point. Secondly, the biorthogonal complete set is employed to
formulate the density matrix of a thermal state. In this situation, we
cannot give a general conclusion for other non-Hermitian systems, which
should be investigated in the future.

Our study shows that quantum criticality in such a non-Hermitian system
appears remarkably\ even at finite temperature. This paves the way for
experimental detection of quantum criticality in complex-parameter plane. In
experiment, there are several ways to realize the Hermitian Ising model,
such as using trapped ions \cite{K,Porras,Wang,Gong}, an array of cavities
\cite{Joshi,Bardyn}, atoms within a cavity \cite{Parkins}, and Rydberg atoms
in an optical lattice \cite{Tony1,K1,Bason,Dudin}. Recently, it was pointed
that \cite{Tony2}, an imaginary transverse field may be implemented by
optically pumping a qubit state into the auxiliary state.

\appendix*

\section{Exact solution of the $\mathcal{H}$}

\label{sec_appendix}

We consider the solution of the non-Hermitian Hamiltonian of Eq. (\ref%
{H_general}). We start by taking the Jordan-Wigner transformation \cite%
{P.Jordan}%
\begin{eqnarray}
\sigma _{j}^{+} &=&\prod\limits_{l<j}\left( 1-2c_{l}^{\dag }c_{l}\right)
c_{j}, \\
\sigma _{j}^{x} &=&1-2c_{j}^{\dag }c_{j}, \\
\sigma _{j}^{z} &=&-\prod\limits_{l<j}\left( 1-2c_{l}^{\dagger }c_{l}\right)
\left( c_{j}+c_{j}^{\dagger }\right) ,
\end{eqnarray}%
to replace the Pauli operators by the fermionic operators $c_{j}$. Likewise,
the parity of the number of fermions%
\begin{equation}
\Pi =\prod_{l=1}^{2N}\left( \sigma _{l}^{x}\right) =\left( -1\right) ^{N_{p}}
\end{equation}%
\bigskip is a conservative quantity, i.e., $\left[ \mathcal{H},\Pi \right]
=0 $, where $N_{p}=\sum_{j=1}^{2N}c_{j}^{\dag }c_{j}$. Then the Hamiltonian (%
\ref{H_general}) can be rewritten as%
\begin{equation}
\mathcal{H}=\sum_{\sigma =+,-}P_{\sigma }\mathcal{H}_{\sigma }P_{\sigma },
\end{equation}%
where
\begin{equation}
P_{\sigma }=\frac{1}{2}\left( 1+\sigma \Pi \right)
\end{equation}%
is the projector on the subspaces with even ($\sigma =+$) and odd ($\sigma
=- $) $N_{p}$. The Hamiltonian in each invariant subspace has the form%
\begin{eqnarray}
\mathcal{H}_{\sigma } &=&-J\sum\limits_{j=1}^{2N-1}\left( c_{j}^{\dag
}c_{j+1}+c_{j+1}^{\dag }c_{j}+c_{j}^{\dag }c_{j+1}^{\dag
}+c_{j+1}c_{j}\right)  \notag \\
&&+J\sigma \left( c_{2N}^{\dag }c_{1}+c_{1}^{\dag }c_{2N}+c_{2N}^{\dag
}c_{1}^{\dag }+c_{1}c_{2N}\right)  \notag \\
&&-Jg_{j}\sum\limits_{j=1}^{2N-1}\left( 1-2c_{j}^{\dagger }c_{j}\right)
\label{H_sub}
\end{eqnarray}%
taking the Fourier transformation%
\begin{equation}
c_{j}=\frac{1}{\sqrt{N}}\sum\limits_{k_{\zeta }}e^{ik_{\zeta }j}\left\{
\begin{array}{cc}
\alpha _{k_{\zeta }}, & \text{even }j \\
\beta _{k_{\zeta }}, & \text{odd }j%
\end{array}%
\right. ,
\end{equation}%
for the Hamiltonians $\mathcal{H}_{\sigma }$, we have%
\begin{eqnarray}
\mathcal{H}_{\sigma } &=&-J\sum_{k_{\sigma }}H_{k_{\sigma }} \\
H_{k_{\sigma }} &=&\left( e^{ik_{\sigma }}+1\right) \alpha _{k_{\sigma
}}^{\dagger }\beta _{k_{\sigma }}+\left( e^{ik_{\sigma }}-1\right) \alpha
_{k_{\sigma }}^{\dagger }\beta _{-k_{\sigma }}^{\dagger }+\mathrm{H.c.}
\notag \\
&&+2\eta -2\left( \eta +i\xi \right) \alpha _{k_{\sigma }}^{\dagger }\alpha
_{k_{\sigma }}  \notag \\
&&-2\left( \eta -i\xi \right) \beta _{k_{\sigma }}^{\dag }\beta _{k_{\sigma
}}
\end{eqnarray}%
where the momentum $k_{\sigma }$ are defined as $k_{+}=2\left( m+1/2\right)
\pi /N$, $k_{-}=2m\pi /N$, $m=0,1,2,...,N-1$, respectively.

In this paper, we focus on the case with large $N$, in which the effect of $%
k_{+}-k_{-}$\ can be neglected. In the following we will neglect the
subscript $\sigma $\ in $k_{\sigma }$. In order to obtain the eigen states
in both subspaces with even and odd $N_{p}$, what we need is to diagonalize
the Hamiltonian

\begin{equation}
h_{k}=H_{k}+H_{-k}
\end{equation}%
with%
\begin{eqnarray}
H_{k} &=&2e^{ik/2}\cos \left( k/2\right) \alpha _{k}^{\dagger }\beta
_{k}+i2e^{ik/2}\sin \left( k/2\right) \alpha _{k}^{\dagger }\beta
_{-k}^{\dagger }+\mathrm{H.c.}  \notag \\
&&+2\eta -2\left( \eta +i\xi \right) \alpha _{k}^{\dagger }\alpha
_{k}-2\left( \eta -i\xi \right) \beta _{k}^{\dag }\beta _{k},
\end{eqnarray}%
due to the fact $\left[ h_{k},h_{k^{\prime }}\right] \propto 0$. This
procedure is similar to introduce the Bogoliubov transformation for solving
a simple transverse field Ising model. There are two $8$-dimensional
invariant spaces for $h_{k}$, spanned by two sets of bases $\{\alpha _{\pm
k}^{\dagger }\left\vert 0\right\rangle $, $\beta _{\pm k}^{\dagger
}\left\vert 0\right\rangle $, $\alpha _{\pm k}^{\dagger }\alpha _{\mp
k}^{\dagger }\beta _{\pm k}^{\dagger }\left\vert 0\right\rangle $, $\beta
_{\pm k}^{\dagger }\beta _{\mp k}^{\dagger }\alpha _{\pm k}^{\dagger
}\left\vert 0\right\rangle \}$ and $\{\left\vert 0\right\rangle $, $\alpha
_{\pm k}^{\dagger }\beta _{\mp k}^{\dagger }\left\vert 0\right\rangle $, $%
\alpha _{\pm k}^{\dagger }\beta _{\pm k}^{\dagger }\left\vert 0\right\rangle
$, $\alpha _{k}^{\dagger }\alpha _{-k}^{\dagger }\left\vert 0\right\rangle $%
, $\beta _{k}^{\dagger }\beta _{-k}^{\dagger }\left\vert 0\right\rangle $, $%
\alpha _{k}^{\dagger }\beta _{k}^{\dagger }\alpha _{-k}^{\dagger }\beta
_{-k}^{\dagger }\left\vert 0\right\rangle \}$, involving odd and even
numbers of fermions, respectively. Here $\left\vert 0\right\rangle $\ is the
vacuum of fermion operator $c_{j}$,\ i.e., $c_{j}\left\vert 0\right\rangle
=0 $. Fortunately, such two $8\times 8$\ matrices can be diagonalized
analytically in the form%
\begin{equation}
h_{k}\overline{\left\vert n\right\rangle }_{k}=2\epsilon ^{n}\left( k\right)
\overline{\left\vert n\right\rangle }_{k},n\in \left[ 1,16\right] ,
\end{equation}%
where $\overline{\left\vert n\right\rangle }_{k}$\ and $\epsilon ^{n}\left(
k\right) $\ are listed in Eqs. (4)-(10), which can be used to construct the
eigenstates and spectrum of the original Hamiltonian. By the same procedure,
we can obtain the eigenstates of $\mathcal{H}^{\dag }$.

\acknowledgments We acknowledge the support of the National Basic Research
Program (973 Program) of China under Grant No. 2012CB921900 and CNSF (Grant
No. 11374163).

\end{document}